\documentclass[10pt,a4paper,twocolumn]{article}

\usepackage{amsmath,amssymb,graphicx,epsfig}

\usepackage{color}
\definecolor{gray}{rgb}{0.5,0.5,0.5}
\definecolor{brick}{rgb}{0.53,0,0}

\title{\bf How to defuse Earth impact threat announcements}
\author{Germano D'Abramo}

\date{}

\begin{document}

\maketitle

\begin{figure*}[t]
\begin{center}
\includegraphics[height=3.5cm]{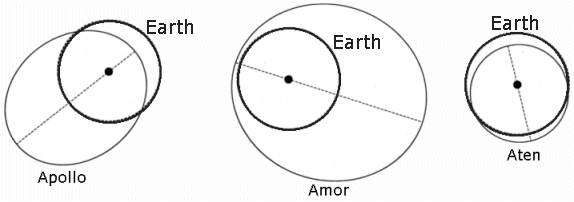}
\end{center}
\caption{
Near Earth Asteroids, or NEAs, are all those asteroids with perihelion 
distance $q$ less than 1.3~A.U. ($1$~A.U. is the mean Earth-Sun distance 
and it is nearly equal to $1.5\times 10^{8}\,$km). The {\em perihelion} 
is the point in the orbit of a planetary object where it is nearest to 
the Sun, as opposed to {\em aphelion} where it is farthest. NEAs are 
divided in three sub-classes: Apollos, Amors, Atens. Apollos have 
$q<0.983\,$A.U.~and semi-major axis $a>1\,$A.U., Amors have 
$1.017\,$A.U.$\,<q<1.3\,$A.U. and Atens have aphelion distance 
$Q>0.983\,$A.U.~and semi-major axis $a<1\,$A.U. Note that 0.983~A.U.~and 
1.017~A.U. are the perihelion and aphelion distances of the Earth, 
respectively. The above picture shows a projection of NEA orbits onto 
the ecliptic plane (the Earth's orbit plane). That could give the wrong 
impression that actually {\em all} Apollo and Aten orbits intersect the 
orbit of the Earth. Obviously, this is not true since orbits are also 
inclined with respect to the ecliptic plane, and rotated around the 
focus (the Sun).}
\label{fig1}
\end{figure*}

\noindent Starting in the 1980s, announcements of newly discovered near 
earth asteroids with unusually high Earth-collision chances in the near 
future have been regularly highlighted in the press, TV and recently on 
the Web. These planetary objects, referred throughout as NEAs, are 
formally defined in Fig.~\ref{fig1}.

Over time, less concern has been noted, unless the announcement concerns 
a new, relatively big asteroid (hundreds of meters in size) with an 
alarming high probability of hitting the Earth in the next decades.

Impact monitoring and analysis is primarily undertaken by the University 
of Pisa (the system is called CLOMON2\footnote{{\it 
http://newton.dm.unipi.it/neodys/index.php?pc=4.1}}) and NASA--JPL 
(National Aeronautics and Space Administration Jet Propulsion 
Laboratory) where it is called SENTRY\footnote{{\it 
http://neo.jpl.nasa.gov/risk/}}. On a daily basis, these two groups 
collect all astrometric data of all NEAs, new and known, observed the 
previous night by the observing surveys.

Astrometric data, or astrometric observations, are the optical/radar 
measurements of the asteroid sky position with respect to the so-called 
``fixed stars" (the Earth, in the case of radar observations). These 
data are compiled from all the observatories around the world and are 
made available at the Minor Planet Center (MPC)\footnote{{\it 
http://www.minorplanetcenter.org/iau/mpc.html}. The MPC designates minor 
bodies in the Solar System and has international responsibility for the 
efficient collection, computation, checking and dissemination of 
astrometric observations and orbits for minor planets and comets.} at 
Harvard University.

SENTRY and CLOMON2 apply their algorithmic procedures daily in order to 
identify possible threatening asteroids among the new ones and to refine 
the orbit of those already categorized as threatening.

The first well-publicized scare concerned asteroid 1997~XF$_{11}$, for 
which a possible impact with the Earth was predicted in 2028 by Brian 
Marsden at MPC. This object and its heralded non-zero impact probability 
in the near future led to unprecedented turmoil within the astronomical 
community and, to a somewhat lesser extent, within the wider community. 
(Note that as described by Chapman (1999), Marsden's initial 
calculations were actually incorrect even given the information known to 
him at the time).

The case of asteroid 1999~AN$_{10}$ was one of the first cases in the 
impact monitoring era (that of CLOMON2 and SENTRY), and this object 
played an important role in developing the concept of \emph{Virtual 
Impactors} that is described below.

One of the most recent asteroids discovered with near future non-zero 
impact probabilities is the $\sim 40\,$ meters sized 
2012~DA$_{14}$, while the most famous one is asteroid 
(99942)~Apophis ($\sim 300\,$~meters in size), which is named after an 
ancient Egyptian evil god. In December, 2004 monitoring systems 
calculated an initial impact chance for the year 2029 as high as $\sim 
1/38$. Then, after more observational data were gathered, the impact in 
2029 was ruled out. Currently, Apophis has a chance nearly equal to 
$\sim 5\times 10^{-6}$ of hitting the Earth in 2068.

For those wanting the most up to date information, the CLOMON2 and 
SENTRY Web pages provide a daily updated list of all potential future 
impactors.

The discovery of new NEAs with non-zero impact probabilities always 
grabs the attention of the astronomical community with a focus on Earth 
impact, as well as the general public when the calculated impact 
probabilities for a specific date in the future result to be {\em 
unusually high} according to the standards of the astronomical community 
(see the International Astronomical Union --IAU-- rules\footnote{\it 
http://web.mit.edu/rpb/wgneo/TechComm.html}).

\begin{figure*}[t]
\begin{center}
\begin{center}
\includegraphics[height=9cm]{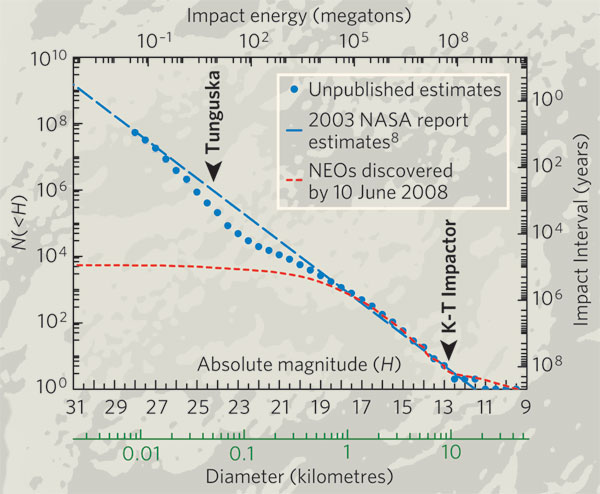}
\end{center}
\end{center}
\caption{
Estimate of the cumulative population of near-Earth asteroids (NEAs) 
versus absolute magnitude $H$, namely brightness at standard distance of 
1 astronomical unit from Earth and Sun (blue-dashed line and blue dots). 
$N(<H)$ is the cumulative number of objects with $H$ less than a given 
value. The fraction currently detected (red dashed line) is nearly 
complete to $H\sim 16$, but falls off rapidly with increasing $H$ 
magnitude, since smaller objects are harder to detect. There are also 
equivalent scales for diameter in km (bottom axis) and expected impact 
interval in years (on the right). The dots represent some unpublished 
estimates of the NEA population. The straight line is a simple power law 
that approximates the estimates: with respect to the right hand vertical 
scale, this line represents $\frac{1}{\rho_i(\geq D)}$, where 
$\rho_i(\geq D)$ is the equation~(\ref{eq0}) in the text. 
Reprinted by permission from Macmillan Publishers Ltd [{\em 
Nature}]: Harris, A.W., 2008: What Spaceguard did. {\em Nature}, {\bf 
453}, 1178-1179. Copyright 2008.}
\label{fig2} 
\end{figure*}

{\em Unusually high} means that the impact probability calculated by 
monitoring systems for a particular new asteroid is higher than the 
yearly statistical impact probability on the Earth coming from an 
unspecified asteroid of equal size or larger belonging to the whole NEA 
population, multiplied by the time interval (in years) which separates 
the present time from the future impact date of the asteroid under 
analysis. This is motivated by the fact that our planet moves in an 
environment swept by the NEA population.

This metric flows from the philosophy of the Palermo Scale, the hazard 
scale most used among astronomers. Using this measure, an impact 
prediction (time, size and probability) is normalized to the {\em 
background expected impact probability} from the present time to the 
time of the prediction.

Expressed as a logarithm, Palermo Scale $0.0$ means the predicted event 
is essentially ``as expected'' in that time interval; positive values 
mean the predicted event is ``extraordinary'' and deserves attention; 
and negative values indicate the predicted event is only a small 
addition to the expected impact flux (see Chesley {\em et al.}, 2002).

Astronomers first estimate the ``per object'' mean impact frequency. 
This is done by generating a synthetic population of point-like objects 
which is thought to be representative of the overall orbital 
distribution of the actual NEA population. Next, they numerically 
integrate over their motion and study the distribution of close 
encounters with the Earth. By extrapolating these statistics down to the 
Earth radius and dividing by the number of point-like objects, they 
obtain the ``per object'' impact frequency. The {\em background impact 
probability} is obtained by multiplying the ``per object'' impact 
frequency with the estimated number of NEAs in different size ranges 
(see Morrison~{\em et al.}, 2003). The background impact probability can 
be approximated by the following power function:

\begin{equation}
\rho_i(\geq D)=20D^{-2.4}\,\textrm{yr}^{-1},
\label{eq0}
\end{equation}
where $D$ is the diameter of the asteroid expressed in {\em meters} (see 
Harris, 2008 and Chesley {\it et al.}, 2002). Note that more detailed 
and up-to-date models for impact production rate exist. 

Rigorously speaking, the background impact probability gives the average 
number of NEAs larger than a given size that hit the Earth {\em per 
year}. If this number is less than $0.1$, as happens with bigger 
asteroids, then it expresses a mathematical probability. Otherwise, if 
it is greater than or equal to unity, as happens with (sub)meter-sized 
NEAs that fall on the Earth every year more than once, then it expresses 
a kind of {\em frequency of impacts}. Accordingly, the average number of 
NEAs that hit the Earth {\em per year} with diameters between $D$ and 
$D+\Delta D$ is given by:

\begin{equation}
\rho_i(D+\Delta D; D)=20\bigl(D^{-2.4}-(D+\Delta D)^{-2.4}\bigr)\,
\textrm{yr}^{-1}.
\label{eq01}
\end{equation}

The reciprocal of the background impact probability 
($\frac{1}{\rho_i(\geq D)}$) is the {\em impact interval time} and it 
gives the mean time between two consecutive impacts of asteroids larger 
than a given size (Fig.~\ref{fig2}).

As is clear from Fig.~\ref{fig2}, the constant power law~(\ref{eq0}) is 
only an approximation of various data about NEA size distribution coming 
from different sources: the points plotted in Fig.~\ref{fig2} represent 
unpublished estimates of the NEA population in different size range.

It must be said that every estimate of size distribution (and thus, of 
background impact probability) has its own intrinsic uncertainty which 
is often large and may not be fully characterized. However, this is not 
considered a major concern, since the argument is somewhat independent 
of the background impact probability exact form.

When additional new astrometric observations become available, the 
asteroid orbits are refined and usually the impact possibilities are 
definitively ruled out, as happened with the case of asteroid 
1999AN$_{10}$.

In what follows, we qualitatively describe how the current impact 
monitoring systems compute impact probabilities of newly discovered 
NEAs. Then, we define the {\em a posteriori} conditional probability 
$W$: this is directed related to the background impact probability as 
well as the mean annual frequency with which impact monitoring systems 
find impact threats among newly discovered asteroids. We argue that 
probability $W$ is the appropriate probability measure to help assess 
from the very beginning the future impact chances of asteroids just 
inserted in the risk lists of monitoring systems. We also give an 
estimate for the upper bound of $W$.

\section*{NEA discovery and impact scare}

When a new NEA is discovered by telescopic surveys around the world, a 
preliminary orbit is computed using its positions in the sky over a 
suitable (minimal) interval of time (astrometric observations). Like 
every physical measurement, astrometric ones are affected by errors 
which make the resulting orbit uncertain to some variable degree. 
Sophisticated mathematical and numerical tools are now available to 
allow the propagation of these measurement errors to the six orbital 
elements which identify the orbit of the asteroid (semi-major axis $a$, 
eccentricity $e$, inclination $i$, longitude of ascending node $\Omega$, 
argument of perihelion $\omega$, time of perihelion passage $t$). The 
new NEA, soon after its discovery, is not represented by a single point 
in the 6-dimensional orbital elements space; rather, it is represented 
by an uncertainty region, a 6-dimensional volume with blurred contours. 
Obviously, the volume of this uncertainty region changes (it generally 
shrinks) when additional observations become available and the orbit 
estimate is refined.

When the nominal orbit (which best fits the observations) of the new NEA 
is geometrically close to the orbit of the Earth, and it shares some 
other peculiar orbital characteristics (like encounter timing issues), 
some orbital solutions which lead to a future collision of the asteroid 
with the Earth can not be excluded only on the basis of the available 
astrometric observations. Orbital solutions which lead to a collision 
are inside the uncertainty region and are fully compatible with the 
available astrometric observations and their errors.

In these cases, monitoring systems sample the uncertainty region with an 
appropriate number of sample points according to a suitable 
6-dimensional space distribution closely related to what is currently 
known about error statistics ({\em a priori} uniform or Gaussian). They 
then evaluate the relative probability that the `true' orbit of the 
asteroid is one of the collision ones. Henceforth, we will refer to this 
probability with the symbol ${\cal V}_i$. The collision orbits are 
nowadays commonly called \emph{Virtual Impactors} (or VIs). 
Sometimes ${\cal V}_i$ is also referred to as {\em VI impact 
probability}.

If a newly discovered NEA exhibits VIs, then it is promptly added by 
monitoring systems into their publicly available risk lists, together 
with its estimated probability ${\cal V}_i$ and Palermo Scale rating.

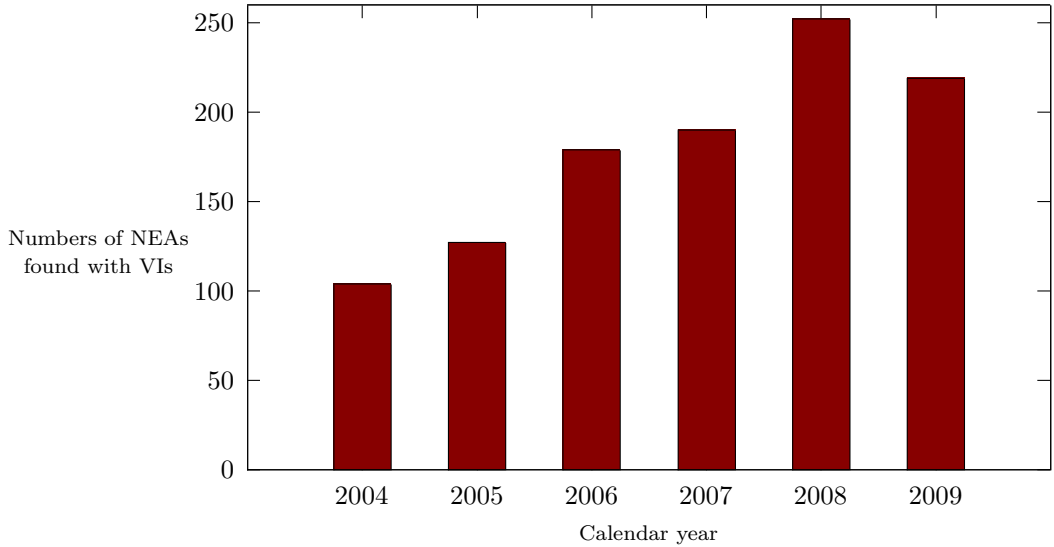
\begin{figure*}[t]
\begin{center}
\setlength{\unitlength}{0.240900pt}
\ifx\plotpoint\undefined\newsavebox{\plotpoint}\fi
\sbox{\plotpoint}{\rule[-0.200pt]{0.400pt}{0.400pt}}%
\begin{picture}(1500,900)(0,0)
\sbox{\plotpoint}{\rule[-0.200pt]{0.400pt}{0.400pt}}%
\put(171.0,131.0){\rule[-0.200pt]{4.818pt}{0.400pt}}
\put(151,131){\makebox(0,0)[r]{ 0}}
\put(1399.0,131.0){\rule[-0.200pt]{4.818pt}{0.400pt}}
\put(171.0,271.0){\rule[-0.200pt]{4.818pt}{0.400pt}}
\put(151,271){\makebox(0,0)[r]{ 50}}
\put(1399.0,271.0){\rule[-0.200pt]{4.818pt}{0.400pt}}
\put(171.0,411.0){\rule[-0.200pt]{4.818pt}{0.400pt}}
\put(151,411){\makebox(0,0)[r]{ 100}}
\put(1399.0,411.0){\rule[-0.200pt]{4.818pt}{0.400pt}}
\put(171.0,551.0){\rule[-0.200pt]{4.818pt}{0.400pt}}
\put(151,551){\makebox(0,0)[r]{ 150}}
\put(1399.0,551.0){\rule[-0.200pt]{4.818pt}{0.400pt}}
\put(171.0,691.0){\rule[-0.200pt]{4.818pt}{0.400pt}}
\put(151,691){\makebox(0,0)[r]{ 200}}
\put(1399.0,691.0){\rule[-0.200pt]{4.818pt}{0.400pt}}
\put(171.0,831.0){\rule[-0.200pt]{4.818pt}{0.400pt}}
\put(151,831){\makebox(0,0)[r]{ 250}}
\put(1399.0,831.0){\rule[-0.200pt]{4.818pt}{0.400pt}}
\put(349.0,131.0){\rule[-0.200pt]{0.400pt}{4.818pt}}
\put(349,90){\makebox(0,0){2004}}
\put(349.0,839.0){\rule[-0.200pt]{0.400pt}{4.818pt}}
\put(528.0,131.0){\rule[-0.200pt]{0.400pt}{4.818pt}}
\put(528,90){\makebox(0,0){2005}}
\put(528.0,839.0){\rule[-0.200pt]{0.400pt}{4.818pt}}
\put(706.0,131.0){\rule[-0.200pt]{0.400pt}{4.818pt}}
\put(706,90){\makebox(0,0){2006}}
\put(706.0,839.0){\rule[-0.200pt]{0.400pt}{4.818pt}}
\put(884.0,131.0){\rule[-0.200pt]{0.400pt}{4.818pt}}
\put(884,90){\makebox(0,0){2007}}
\put(884.0,839.0){\rule[-0.200pt]{0.400pt}{4.818pt}}
\put(1062.0,131.0){\rule[-0.200pt]{0.400pt}{4.818pt}}
\put(1062,90){\makebox(0,0){2008}}
\put(1062.0,839.0){\rule[-0.200pt]{0.400pt}{4.818pt}}
\put(1241.0,131.0){\rule[-0.200pt]{0.400pt}{4.818pt}}
\put(1241,90){\makebox(0,0){2009}}
\put(1241.0,839.0){\rule[-0.200pt]{0.400pt}{4.818pt}}
\put(171.0,131.0){\rule[-0.200pt]{0.400pt}{175.375pt}}
\put(171.0,131.0){\rule[-0.200pt]{300.643pt}{0.400pt}}
\put(1419.0,131.0){\rule[-0.200pt]{0.400pt}{175.375pt}}
\put(171.0,859.0){\rule[-0.200pt]{300.643pt}{0.400pt}}
\put(-60,495){\makebox(0,0){\footnotesize Numbers of NEAs}}
\put(-60,450){\makebox(0,0){\footnotesize found with VIs}}
\put(795,29){\makebox(0,0){\footnotesize Calendar year}}
%
\textcolor{brick}{\put(305,131){\rule{21.681pt}{70.3428pt}}}
\put(305.0,131.0){\rule[-0.200pt]{0.400pt}{70.102pt}}
\put(305.0,422.0){\rule[-0.200pt]{21.440pt}{0.400pt}}
\put(394.0,131.0){\rule[-0.200pt]{0.400pt}{70.102pt}}
\put(305.0,131.0){\rule[-0.200pt]{21.440pt}{0.400pt}}
\textcolor{brick}{\put(483,131){\rule{21.681pt}{86.0013pt}}}
\put(483.0,131.0){\rule[-0.200pt]{0.400pt}{85.760pt}}
\put(483.0,487.0){\rule[-0.200pt]{21.440pt}{0.400pt}}
\put(572.0,131.0){\rule[-0.200pt]{0.400pt}{85.760pt}}
\put(483.0,131.0){\rule[-0.200pt]{21.440pt}{0.400pt}}
\textcolor{brick}{\put(661,131){\rule{21.681pt}{120.932pt}}}
\put(661.0,131.0){\rule[-0.200pt]{0.400pt}{120.691pt}}
\put(661.0,632.0){\rule[-0.200pt]{21.440pt}{0.400pt}}
\put(750.0,131.0){\rule[-0.200pt]{0.400pt}{120.691pt}}
\put(661.0,131.0){\rule[-0.200pt]{21.440pt}{0.400pt}}
\textcolor{brick}{\put(840,131){\rule{21.681pt}{128.4pt}}}
\put(840.0,131.0){\rule[-0.200pt]{0.400pt}{128.159pt}}
\put(840.0,663.0){\rule[-0.200pt]{21.440pt}{0.400pt}}
\put(929.0,131.0){\rule[-0.200pt]{0.400pt}{128.159pt}}
\put(840.0,131.0){\rule[-0.200pt]{21.440pt}{0.400pt}}
\textcolor{brick}{\put(1018,131){\rule{21.681pt}{170.316pt}}}
\put(1018.0,131.0){\rule[-0.200pt]{0.400pt}{170.075pt}}
\put(1018.0,837.0){\rule[-0.200pt]{21.440pt}{0.400pt}}
\put(1107.0,131.0){\rule[-0.200pt]{0.400pt}{170.075pt}}
\put(1018.0,131.0){\rule[-0.200pt]{21.440pt}{0.400pt}}
\textcolor{brick}{\put(1196,131){\rule{21.681pt}{147.913pt}}}
\put(1196.0,131.0){\rule[-0.200pt]{0.400pt}{147.672pt}}
\put(1196.0,744.0){\rule[-0.200pt]{21.440pt}{0.400pt}}
\put(1285.0,131.0){\rule[-0.200pt]{0.400pt}{147.672pt}}
\put(1196.0,131.0){\rule[-0.200pt]{21.440pt}{0.400pt}}
\put(171.0,131.0){\rule[-0.200pt]{0.400pt}{175.375pt}}
\put(171.0,131.0){\rule[-0.200pt]{300.643pt}{0.400pt}}
\put(1419.0,131.0){\rule[-0.200pt]{0.400pt}{175.375pt}}
\put(171.0,859.0){\rule[-0.200pt]{300.643pt}{0.400pt}}
\end{picture}
\end{center}
\caption{
NEAs discovered between calendar years 2004 and 2009 exhibiting VI 
orbital solutions (taken from SENTRY risk list and archive). The 
increasing trend with time is mainly due to the fact that discovery 
surveys have become more efficient in discovering NEAs over the years. 
The efficiency of impact monitoring systems in finding VIs is very high 
and it has been almost always the same through the years.}
\label{fig3}
\end{figure*}

CLOMON2 and SENTRY find hundreds of newly discovered NEAs with VI 
orbital solutions every year (see Fig.~\ref{fig3} for a summary of of VI 
identification between calendar years 2004 and 2009).

Every time additional astrometric observations become available, the 
characterization of the asteroid orbit improves and the estimated impact 
probability ${\cal V}_i$ is re-computed. This may happen in the weeks, 
months and even years following the discovery date. A typical pattern is 
that as the orbit becomes more precisely determined, impact probability 
${\cal V}_i$ often increases initially, but then decreases until it 
falls to zero, or some very low number.

The reason for the initial increasing behavior is rather technical: 
since the uncertainty region generally shrinks with new additional 
observations, some VI orbital solutions often remain inside the 
uncertainty region in the elements space.

In the following section we propose an {\em a posteriori} conditional 
reading of VI impact probabilities. We label our probability as `{\em a 
posteriori}' since it is obtained as the ratio between two statistical 
quantities (relative frequency). On the other hand, the VI impact 
probabilities can be considered {\em a priori}, in the sense that they 
are obtained through sophisticated mathematical models and deductive 
reasoning (for instance, the choice of an {\em a priori} 6-dimensional 
space distribution to sample the uncertainty region).

\section*{Statistical reading of ${\cal V}_i$: the probability $W$}

When a newly discovered NEA is found, a key question is whether the 
probability that ${\cal V}_i$ approaches and eventually reaches unity 
(within this paper we will use the compact notation `${\cal V}_i \to 
1$'), after the right amount of additional new astrometric observations 
has become available. This is equivalent to asking whether {\em only 
knowing} that a newly discovered NEA exhibits some VI orbital solutions, 
what is the probability that ${\cal V}_i$ will be equal to $1$ at the 
end of the whole orbital refinement process?

The following thought experiment helps motivate this point. Suppose that 
the existing discovery surveys are able to discover {\em all} NEAs which 
pass close to the Earth down to a size cut-off. This is obviously not 
true since during their close approaches to the Earth, many unknown 
asteroids remain too dim to be detected by telescopes: they are too 
small in size and/or still `too distant'. Moreover, some NEAs are not 
found because telescopic observations miss them, namely surveys do not 
image a portion of the night sky when they are there and bright enough 
to be seen.  This becomes especially true for asteroids relatively close 
to the Sun, where monitoring is more sporadic or even impossible.

We also suppose that every discovered asteroid really impacting the 
Earth in the future will show some VIs, with low ${\cal V}_i$ soon after 
the discovery and fluctuating with an increasing trend as soon as 
subsequent astrometric observations become available, as usually happens 
in reality.

In other words, we are putting ourselves in the somewhat idealized 
situation where {\em every} impacting asteroid above a size cut-off is 
surely discovered and monitoring systems {\em surely} spot some VIs for 
it soon after its discovery.

Thus, we define $W$ as:

\begin{equation}
W(D+\Delta D;D)= \lim_{T\to\infty}\frac{n(D+\Delta D;D)}{v(D+\Delta D;D)},
\label{eq1a}
\end{equation}
where $n(D+\Delta D;D)$ is the number of asteroids with size between $D$ 
and $D+\Delta D$ which {\em actually} impact the Earth in the period of 
time $T$, with $T\gg 1\,$year, and $v(D+\Delta D;D)$ is the number of 
asteroids found by monitoring systems among all the newly discovered 
ones to exhibit VI orbital solutions, in the same size interval and in 
the same period of time $T$.

The quantity $W$ can be seen as the {\em a posteriori} conditional 
probability of ${\cal V}_i \to 1$, and could also be interpreted as a 
kind of `weight' of the VI impact probability calculation. Now, by 
dividing both numerator and denominator of eq.~(\ref{eq1a}) by $T$, we 
have:
 
\begin{equation*}
W(D+\Delta D;D)=\lim_{T\to\infty}\frac{n(D+\Delta D;D)}{T}
\frac{1}{\frac{v(D+\Delta D;D)}{T}}
\end{equation*}
\begin{equation}
=\frac{\rho_{i}(D+\Delta D;D)}{f_{{\cal V}_i} (D+\Delta D;D)}.
\label{eq1}
\end{equation}

The limit $\lim_{T\to\infty}\frac{n(D+\Delta D;D)}{T}$ is the definition 
of the background annual impact probability, equation~(\ref{eq01}). The 
function $f_{{\cal V}_i}(D+\Delta D;D)= 
\lim_{T\to\infty}\frac{v(D+\Delta D;D)}{T}$ is the annual frequency of 
newly discovered NEAs with sizes between $D$ and $D+\Delta D$ found with 
VI orbital solutions.

Note that, according to the earlier assumption, the number $n(D+\Delta 
D;D)$ is counted in the number $v(D+\Delta D;D)$, if every impacting 
asteroid is identified soon after its discovery as having some VI 
orbital solutions, thus $n(D+\Delta D;D)$ is always less than or equal 
to $v(D+\Delta D;D)$.

We focus our attention on eq.~(\ref{eq1}). Within the hypotheses 
introduced above, we imagine waiting for a long period (many years), 
$T$, and count the number $n(D+\Delta D;D)$ of true asteroid impacts on 
the Earth and the number $v(D+\Delta D;D)$ of newly discovered NEAs 
found by monitoring systems to have VI orbital solutions with size 
between $D$ and $D+\Delta D$ during that period of time. The limit for 
$T\to\infty$ of the ratio between these two numbers is the conditional 
probability that a discovered asteroid of size between $D$ and $D+\Delta 
D$ will {\em eventually} fall on Earth, given that it has VI orbital 
solutions. In eq.~(\ref{eq1}) we have simply rewritten $W$ in terms of 
the background annual impact frequency $\rho_{i}$ and the mean annual VI 
detection frequency $f_{{\cal V}_i}$.

Thus, $W(D+\Delta D;D)$ gives the probability that an asteroid with VIs 
(an asteroid inserted into the risk lists of the monitoring systems) has 
its probability ${\cal V}_i$ eventually reaching unity.

We noted earlier that $W$ can be interpreted as a kind of `weight' of 
the VI impact probability calculation. Suppose that, thanks to 
improvements in observational techniques (e.g.~higher positional 
precision) and orbital computation, the number of newly discovered 
asteroids identified by monitoring systems as potential impactors 
decreases in every diameter (or absolute magnitude) range. Accordingly, 
the probability $W$ will increase (given its definition) in every 
diameter (or absolute magnitude) range. Since the decrease of the number 
of potential impactors among the new asteroids means an increased 
capability in constraining the true potential impactors, the consequent 
increase of $W$ would equivalently mean an increased capability in 
constraining the true potential impactors by the monitoring system. 
Therefore, $W$ could be seen as a `weight' in expressing the actual 
capabilities of the monitoring system.

Moreover, we can see that $W$ is not directly related to the specific 
numerical value of ${\cal V}_i$, no matter how ${\cal V}_i$'s specific, 
fluctuating numerical figure is. Rather, it depends upon $f_{{\cal 
V}_i}$ which, in turn, depends upon observational characteristics. These 
characteristics are the annual number of NEA discoveries, the amount of 
astrometric observations available at discovery, the magnitude of 
astrometric errors (and conventions in their statistical treatment), as 
well as the observational geometry and orbital characteristics of the 
newly discovered asteroids.

\begin{figure*}[t]
\begin{center}
\input{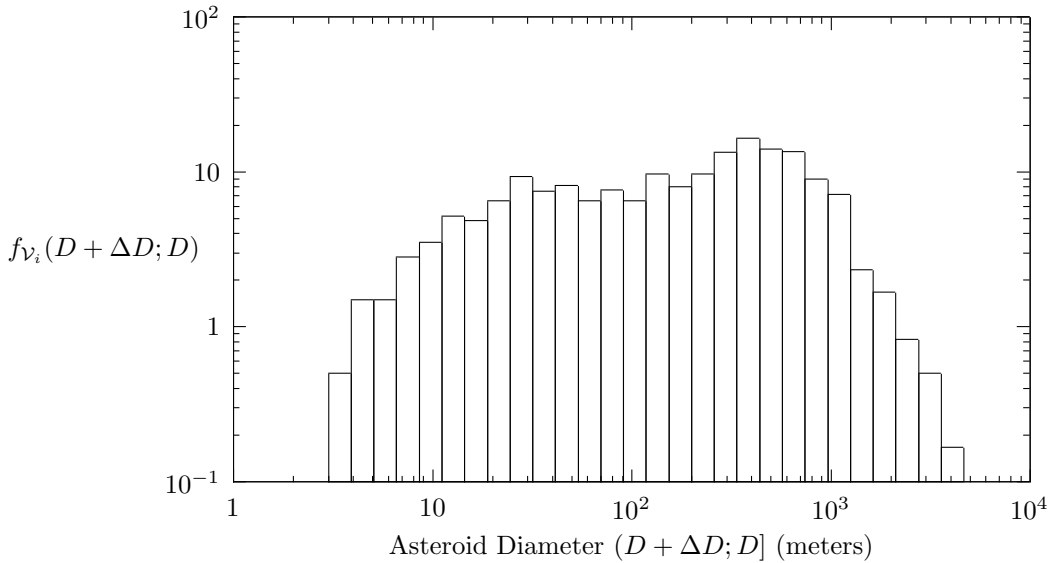}
\end{center}
\caption{
Annual frequency $f_{{\cal V}_i}(D+\Delta;D)$ of VI detections, 
estimated using all NEAs discovered between calendar years 2004 and 
2009 and listed in the SENTRY risk list (and archive). The size range of 
this group is from $\sim 3\,$m up to a maximum of $\sim 4\,$km. The size 
bins are spaced geometrically (such that $D+\Delta D=1.3D$).}
\label{fig4} 
\end{figure*}

\begin{figure*}[t]
\begin{center}
\input{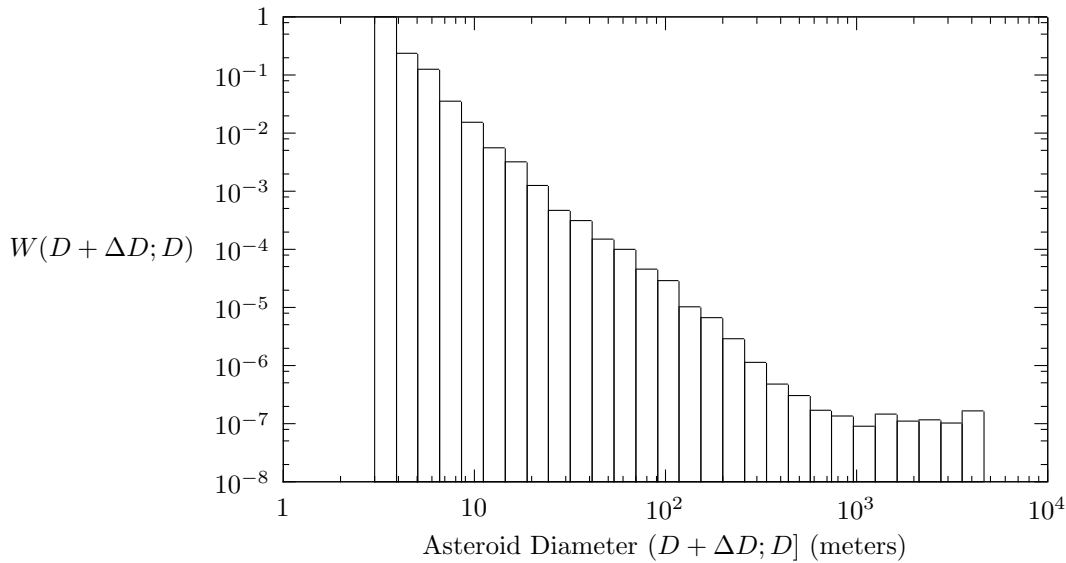}
\end{center}
\caption{
The probability $W(D+\Delta D; D)$ as a function of asteroid diameter. 
The size bins are spaced geometrically (such that $D+\Delta 
D=1.3D$). $W(D+\Delta D;D)$ represents the probability that an asteroid 
with VIs, and thus an asteroid inserted into the risk lists of the 
monitoring systems, has its probability ${\cal V}_i$ eventually reaching 
unity. It is thought that the whole population of NEAs with diameter 
larger than nearly 4~km has probably already been discovered and the 
orbits are also known to be safe for the Earth. This is the reason why 
no VI has been found so far for such NEAs.}
\label{fig5}
\end{figure*}

Although the value of $f_{{\cal V}_i}$ depends upon contingent, variable 
features, it is worthwhile to estimate it statistically. Given the total 
number of NEAs with VIs found at every size between calendar years 2004 
and 2009 (Fig.~\ref{fig3}) an indicative estimate of $f_{{\cal 
V}_i}(D+\Delta D; D)$ is possible (see Fig.~\ref{fig4}). Consequently a 
preliminary estimate of $W(D+\Delta D; D)$ can be obtained as in 
Fig.~\ref{fig5}. The annual VI detection frequency $f_{{\cal 
V}_i}(D+\Delta D; D)$ shown in Fig.~\ref{fig4} has been obtained by 
dividing the number of all NEAs discovered between calendar years 2004 
and 2009, and listed in the SENTRY risk list, by 6 (years) and binning 
the result by size.

Relaxing the optimistic assumptions on the ``almost perfect NEA 
discovery efficiency'' and VI monitoring capabilities makes $f_{{\cal 
V}_i}$, as approximated with the aid of Fig.~\ref{fig4}, even a lower 
limit. As a result, the computation of $W(D+\Delta D; D)$ is surely an 
overestimate.

Furthermore, $f_{{\cal V}_i}$ changes over time, since the discovery 
completion increases and the discovery rate declines with time (because 
the NEA population is stable). It is useful to update $f_{{\cal V}_i}$ 
(and hence $W$) from time to time. But the sense and the validity of the 
definition of $W$ are not affected by (and dependent on) such a time 
dependence.

In summary, the function $W(D+\Delta D;D)$ provides us with a first 
simple tool to evaluate the chance that ${\cal V}_i$ (the impact 
probability calculated by the monitoring systems for newly discovered 
NEAs with VI orbital solutions) eventually reaches unity. As a matter of 
fact, ${\cal V}_i$ is a stochastic variable since nobody knows how 
${\cal V}_i$ will evolve with additional observations, and it is 
perfectly legitimate (and valuable) to define a {\em probability} ($W$) 
of a (stochastic) {\em probability value} (${\cal V}_i$).

Consider now the application of probability measure $W$ to two well 
known cases: that of Apophis as well as asteroid 2008~TC$_3$ (which 
actually impacted the Earth the day after its discovery). These examples 
suggest the reliability of probability $W$ in providing an early direct 
glimpse of the most likely fate of probability ${\cal V}_i$.

Soon after Apophis was discovered in December 2004, impact monitoring 
systems identified multiple VIs orbital solutions and obtained the 
fear-inducing initial impact probability of $\sim 1/38$ for the year 
2029.

What is the probability $W$ for Apophis, an asteroid with an estimated 
diameter of nearly $300\,$m? According to Figure~\ref{fig5}, $W(\sim 
300\,\textrm{m})$ is less than $10^{-6}$, namely more than four orders 
of magnitude lower than that initially reported by monitoring systems 
and close to their current estimate, obtained after some orbit 
refinement.

Nowadays, it is almost certain that the probability ${\cal V}_i$ for 
Apophis will go to zero with future astrometric observations, but it was 
not so clear at the beginning of the impact monitoring process (with an 
initial ${\cal V}_i$ of $\sim 1/38$). As a matter of fact, an early use 
of probability $W$ would have given a direct glimpse of what would have 
been the most likely fate of ${\cal V}_i$ for Apophis.

The same would have happened for 2008~TC$_3$, which was discovered on 
October 6, 2008 and impacted the Earth about 20 hours later. The impact 
was predicted by monitoring systems to have a probability of $\sim 
100\%$. Consistently, the probability $W$ for objects in the size range 
of 2008~TC$_3$ is practically 1, the size of that asteroid being 
estimated to be between $2$ and $4\,$m (see Fig~\ref{fig5}).

In the end, the proper impact probability of a newly discovered asteroid 
is not ${\cal V}_i$ (which actually fluctuates with new additional 
observations) but the probability that ${\cal V}_i\to 1$, i.~e. $W$.

\section*{Further Reading}

\noindent Chapman, C.R. 1999. The asteroid/comet impact hazard. Case 
Study for {\it Workshop on Prediction in the Earth Sciences: Use and 
Misuse in Policy Making}, July 10-12 1997 - Natl.~Center for Atmospheric 
Research, Boulder, CO and September 10-12 1998, Estes Park, CO. 
Available on-line at: {\it www.boulder.swri.edu/clark/ncar799.html}\\

\noindent Chesley, R.S., Chodas, P.W., Milani, A., Valsecchi, G.B., 
Yeomans, D.K. 2002. Quantifying the risk posed by potential Earth 
impacts. {\it Icarus} {\bf 159}, 423.\\

\noindent Harris, A.W. 2008. What Spaceguard did. {\it Nature} {\bf 
453}, 1178.\\

\noindent Milani, A., Chesley, S.R., Chodas, P.W., Valsecchi, G.B. 
Asteroid close approaches and impact opportunities. In: W.~Bottke, 
A.~Cellino, P.~Paolicchi and R.P.~Binzel, Editors.  Asteroids III. 
Tucson: University of Arizona Press; 2003.\\

\noindent Morrison, D., Harris, A.W., Sommer, G., Chapman, C.R., Carusi, 
A. Dealing with the Impact Hazard. In: W.~Bottke, A.~Cellino, 
P.~Paolicchi and R.P.~Binzel, Editors. Asteroids III. Tucson: University 
of Arizona Press; 2003.

\end{document}